# Effect of E-printing on Citation Rates in Astronomy and Physics

By Edwin A. Henneken, Michael J. Kurtz, Guenther Eichhorn, Alberto Accomazzi, Carolyn Grant, Donna Thompson, and Stephen S. Murray

Harvard-Smithsonian Center for Astrophysics
Cambridge, MA 02138, USA

point of contact: ehenneken@cfa.harvard.edu

**Abstract**

In this report we examine the change in citation behavior since the introduction of the arXiv e-print repository (Ginsparg, 2001). It has been observed that papers that initially appear as arXiv e-prints get cited more than papers that do not (Lawrence, 2001; Brody et al., 2004; Schwarz & Kennicutt, 2004; Kurtz et al., 2005a, Metcalfe, 2005). Using the citation statistics from the NASA-Smithsonian Astrophysics Data System (ADS; Kurtz et al., 1993, 2000), we confirm the findings from other studies, we examine the average citation rate to e-printed papers in the Astrophysical Journal, and we show that for a number of major astronomy and physics journals the most important papers are submitted to the arXiv e-print repository first.

In the 1990s publishers and scholars began to take note of the effect of early access and open access to scholarly papers because the Internet had made both possible on a grand scale. The two approaches are subtly different. Early access is availability of a paper prior to its publication in a journal in the form of an e-print (electronic preprint); open access is free, unrestricted availability of papers (examples are freely available e-prints, freely available conference proceedings and electronic journals that offer their papers without a subscription fee). A wide range of studies has been published discussing the influence of early and open availability of scholarly papers on their subsequent accumulation of citations. It has been shown that papers available through open access sources (in particular the arXiv e-print repository) are more heavily cited than papers that are not (Lawrence, 2001; Brody et al., 2004). Kurtz et al. (2005a) examines this effect. They claim that there are (at least) three possible, non-exclusive effects that cause this effect of higher citation rates. The paper formulates these three effects in three postulates and then continues to test these postulates on data taken from the Astrophysics Data System and the arXiv. These postulates are: Open Access, Early Access and Self-selection Bias. The postulates are defined as follows (see Kurtz et al., 2005a):

> **Open Access Postulate (OA):** *Because of free, unrestricted access, papers are read more easily and therefore get cited more frequently*
> **Early Access Postulate (EA):** *Papers offered as e-print are available sooner and therefore gain primacy and additional time in press, and therefore they get cited more often*
> **Self-selection Bias Postulate (SB):** *The most important -- and therefore most citable -- papers are posted on the Internet*

During the 1990s both the Astrophysics Data System (Kurtz et al., 1993) and the arXiv (Ginsparg, 2001) were founded. These events improved the access to scholarly literature dramatically. Between 1996 and 1999 freely available, digitized versions of all back issues of the seven core astronomy journals became available on line. This free access to back issues is an example of pure open access, and allows us to see the Kurtz open-access postulate in action, while the free access to e-prints via the arXiv, is an example both open access and early access, and allows us to see of the combined effect of the two postulates.

The study by Kurtz et al. (2005a) tests the OA and EA postulates by looking at the probability that a given paper cites another paper published during a particular period, based on an algorithm that Kurtz et al. created. Using this algorithm, they found no increase in citation rates of older papers as a result of the free availability of back issues, but a signification increase in citation rates for recent papers. So it is not open access that affects citations, but early access – a theory they said was supported by the fact that astronomers have no problem getting access to papers. "Because the marginal cost of being an astronomer with access to the core literature is so much lower than the cost of being an astronomer in the first place, it is reasonable to postulate that essentially all astronomers have access to the core literature."

Using various statistical analyses, the Kurtz group was able to identify the additional effect of their SB (self-selection bias) postulate. They found that there are significantly fewer papers that have not appeared as e-prints in the top 200 most cited papers than they would have expected based only on their OA and EA postulates. Their conclusion was that "because papers in the arXiv are not refereed ... this suggests that authors self-censor or self-promote, or that for some reason the most citable authors are also those who first use the new publication venue".

**Building on Kurtz et al.**

Our report focuses on fact finding, and through fact finding we support the findings by the Kurtz group. We concentrate on showing that there is a significant difference between papers that appear as arXiv e-prints and papers that do not -- the amount of citations received by papers in each group. We limit our analysis to four major astronomy journals (The Astrophysical Journal, including Letters and Supplement Series, The Monthly Notices of the Royal Astronomical Society, Astronomy & Astrophysics, Proceedings of the Astronomical Society of the Pacific) and two physics journals (Physical Review and Nuclear Physics). We found a major difference between the normalized citation rate for papers from the pre-arXiv era and papers that have been offered as e-prints in the arXiv repository. Furthermore, based on a dataset of the Astrophysical Journal in the periods of 1985 through 1987 and 1997 through 1999, we follow the citations for traditionally published papers and e-printed papers.

**2. Data**

We base our analysis on citation data in the Astrophysics Data System database, which are nearly 100% complete for the core astronomy journals. We cannot claim to be as complete for the physics journals, but the citation data for the physics journals used in our report is complete enough to contain a truthful representation of the trends we are investigating, which are the same in astronomy and physics. The citation data in the ADS database are obtained by resolving the references in papers, when available, and by user submission. The process of reference "resolving" means parsing publication source data from a reference string and attempting to associate these with a bibliographic record already present in the ADS (see Accomazzi et al., 1999).

As part of the ADS, we create a mapping or concordance between the arXiv e-print identifier and the identifier with which the published journal paper is known within the ADS (the "bibcode"). This process of bibcode matching uses the metadata in the entries on the arXiv (title, author names, submission information) to match the e-print with an existing ADS entry. Since in our report we look at e-prints that have been published in journals (with a granularity of one month), this concordance should ideally represent a 100% match for our dataset. Unfortunately, titles and sometimes author lists have changed by the time a paper is published in a journal. From spot checks and from reviewing the list of the 200 most cited papers we estimate that in reality the concordance achieved by the ADS is around 98% (see Kurtz et al., 2005a). This amount of incompleteness seems to have no influence on any of the outcomes of our findings, however.

**3. Results**

**3.1 The evolution of e-printing**

While journal policies and practices on e-printing may vary by discipline (see e.g. Schwarz & Kennicutt, 2004), for our purposes we merely looked at overall trends within astronomy and physics. Figure 1a shows the percentage of e-printed papers in a number of major astronomy journals from 1992 through 2004. Figure 1b shows it for a number of physics journals (see Appendix A for the journal abbreviations used).

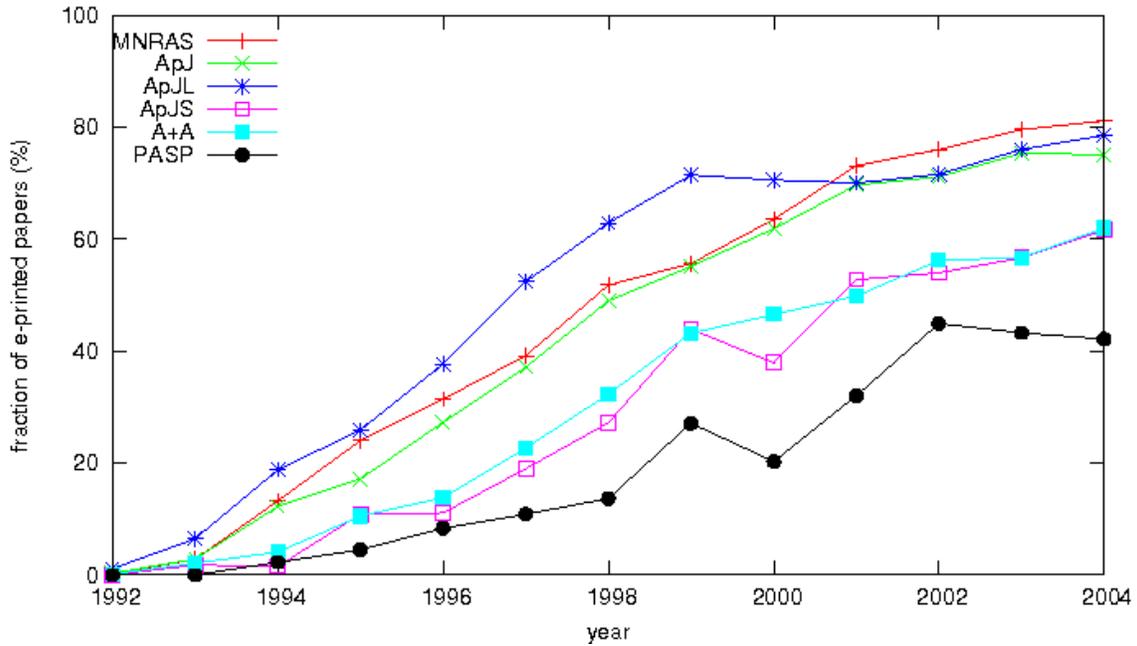

Figure 1a. Percentage of e-printed papers in a number of astronomy journals

With the exception of Physics Review D and Nuclear Physics B (PhRvD and NuPhB), the percentages show a gentle, yet steady growth over time. The picture is more dramatic in most cases when we sort the publications for each year according to citations and take the top 100 of this list.

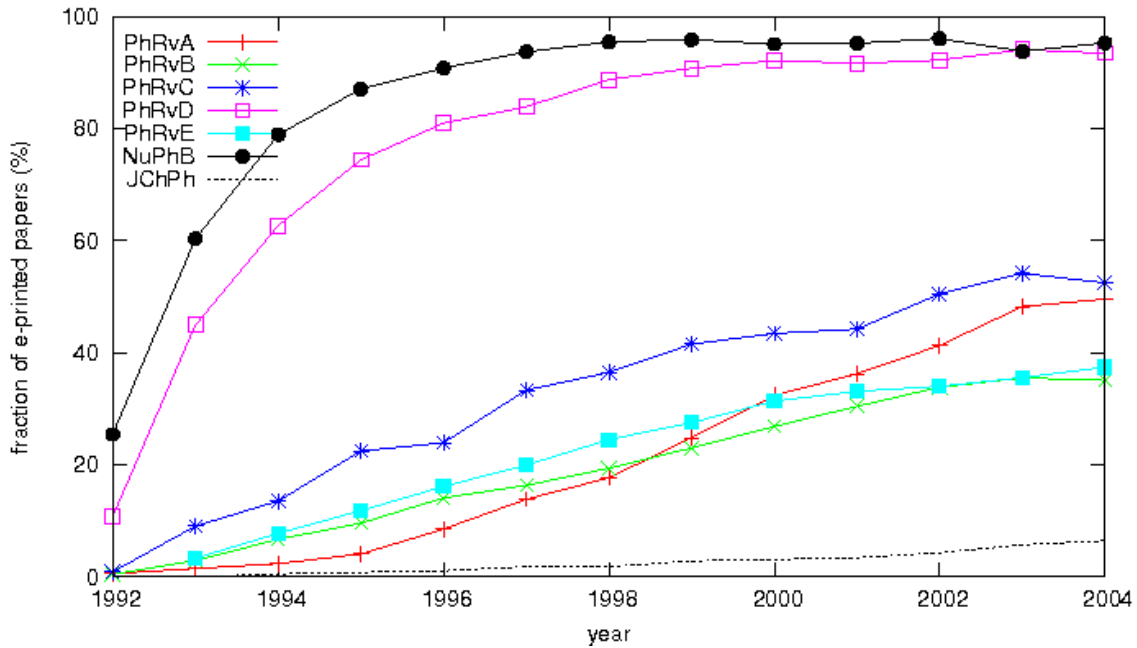

Figure 1b. Fraction of e-printed papers in a number of physics journals

Figure 2 shows the results for this top 100. We see that in astronomy, with exception of Publications of the Astronomical Society of the Pacific (PASP) and The Astrophysical Journal Supplement Series (ApJS), over 90% of the papers of the top 100 have been e-printed, and the figure is close to 100% for Monthly Notices of the Royal Astronomical Society (MNRAS). In the case of physics we see that the Journal of Chemical Physics (JchPh) does not have a significant e-printing culture, meaning that very few papers in JChPh are submitted to the arXiv. It falls outside the scope of this paper to investigate reasons behind this phenomenon.

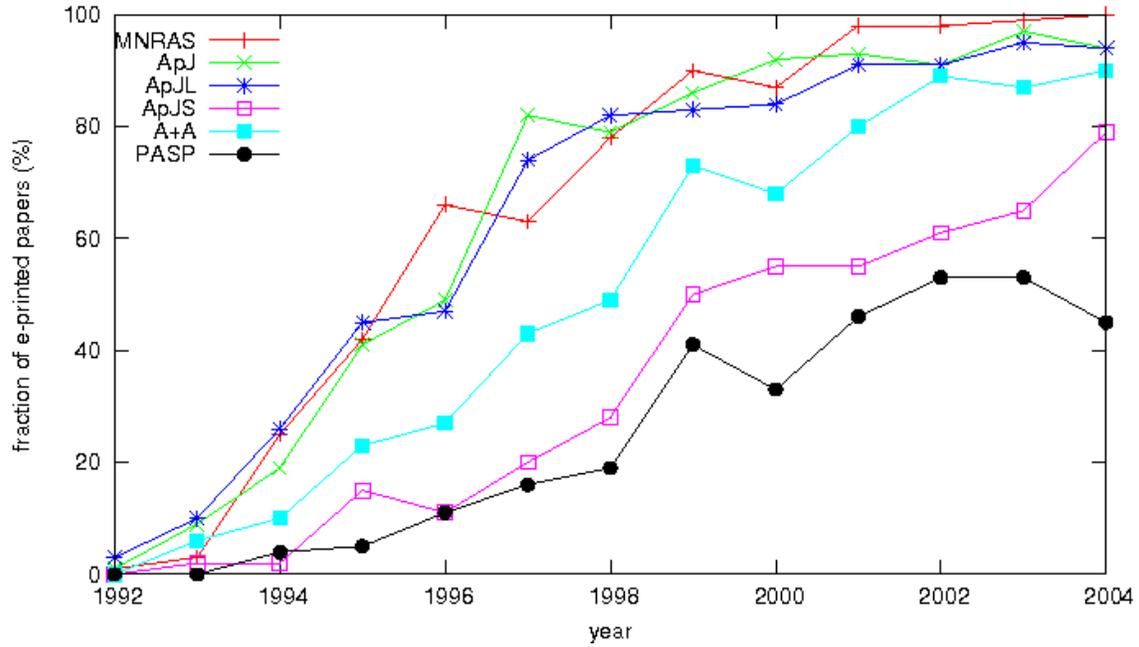

Figure 2 a. Percentage of e-printed papers for the top 100 most cited papers in astronomy

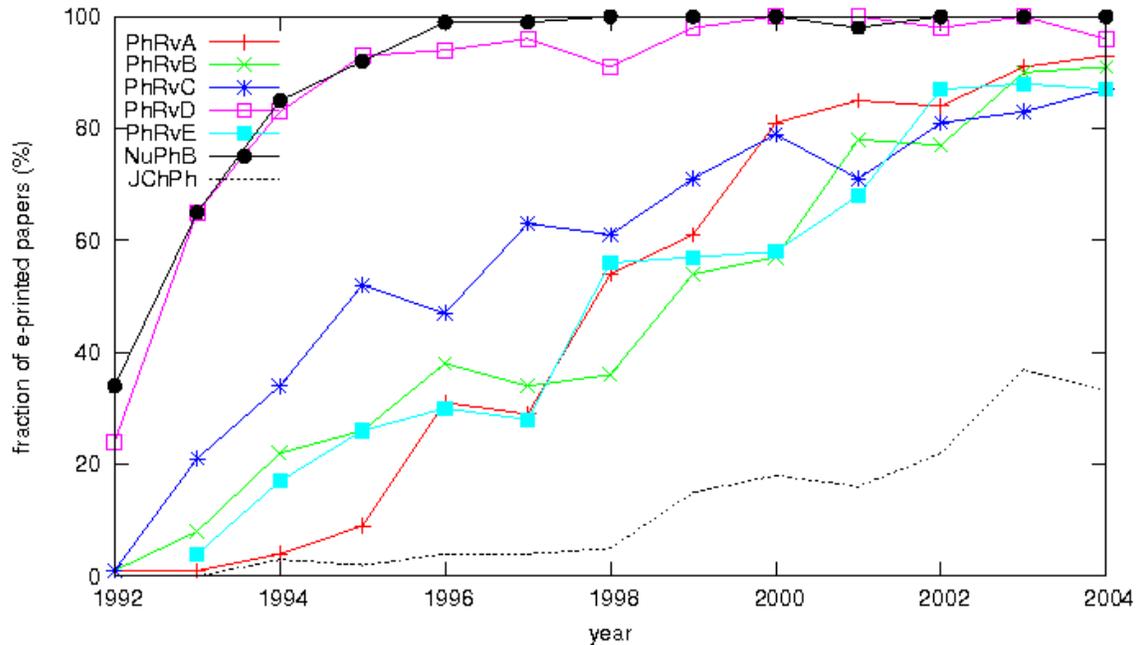

Figure 2 b. Percentage of e-printed papers for the top 100 most cited papers in physics

### 3.2 Impact of e-printing on citation behavior

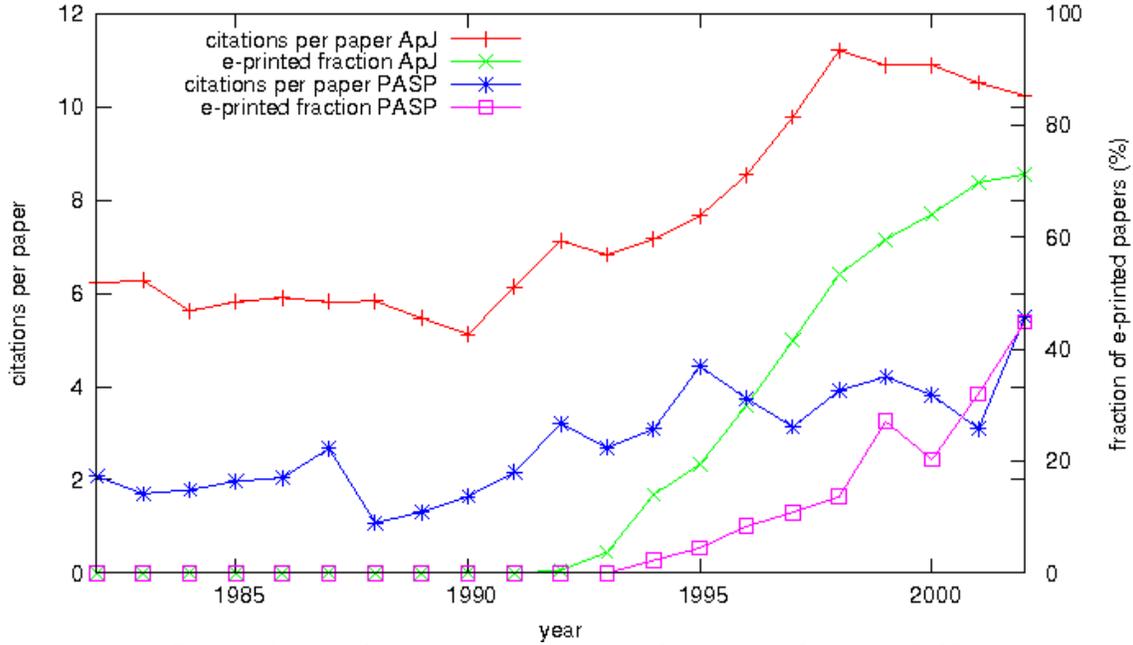

Figure 3a. Number of citations per paper two years after publication for ApJ and PASP

A number of studies show that e-printed papers get cited more often than papers that have not been e-printed. We illustrate this phenomenon using slightly different statistics. In our first example we determine the number of citations per paper two years after the paper has been published. Citations from e-prints are not contributing to these statistics; we follow only citations published in papers in journals. We detrmine this number for papers published in the period of 1982 through 2002. Specifically, we will determine this number for papers published in Astrophysical Journal (ApJ) and Publications of the Astronomical Society of the Pacific (PASP). This comparison is also made for papers published in Physics Review D (PhRvD) and in Physics Review B (PhRvB). The results are shown in figures 3a and 3b, respectively.

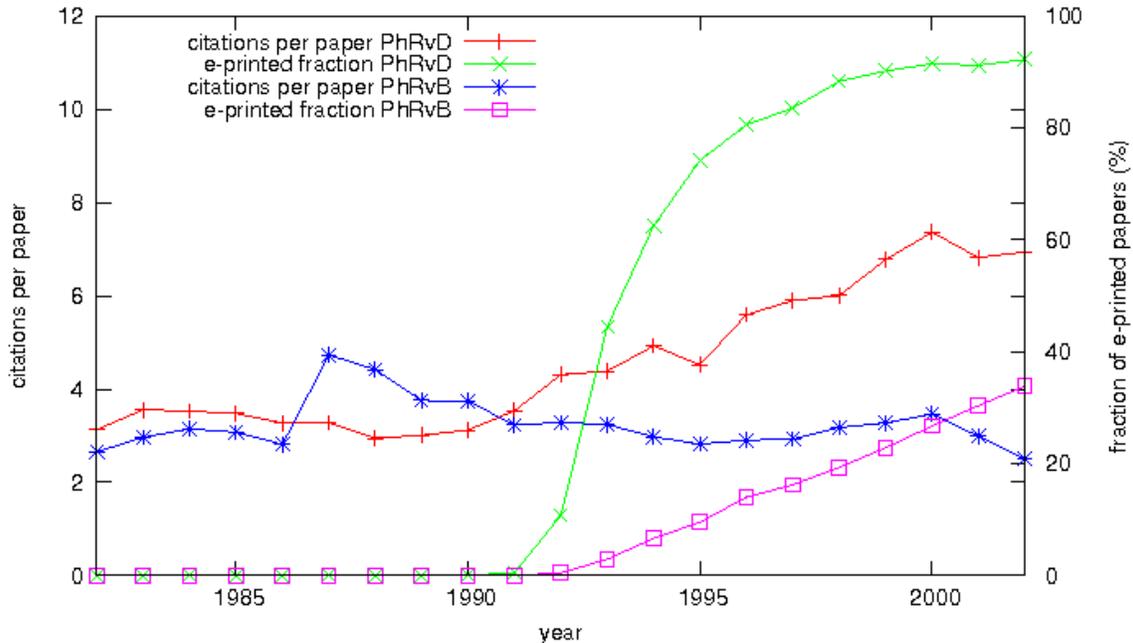

Figure 3b. Number of citations per paper two years after publication for PhRvB and PhRvD

Figure 3a shows that with the introduction of the arXiv e-print repository in 1992, the number of citations per paper after two years starts to increase for ApJ. Although the number increases slightly for PASP as well, it is not as drastic as for ApJ. The reason is probably a combination of factors: subject matter (this journal has more emphasis on instrumentation and catalogs), a lower fraction of e-printed papers and a smaller audience. Other journals contain more papers on trendy, "hot" topics , which on average get more citations.

Figure 3b shows that the number of citations per paper after two years starts to increase for PhRvD, while for PhRvB this number stays rather constant.

Another way of looking at the impact of e-printing is to select a period of time and determine the number of citations after publication as an ensemble average. For our report, we have taken all ApJ papers published from 1985 through 1987, and 1997 through 1999. We follow the citations to these papers for five years after publication of the paper. The period of 1985 through 1987 has been included as a comparison; it is in the pre-arXiv era. We determined how many citations an e-printed paper acquires on average as a function of time after publication, normalized with respect to the mean number of citations over the period of five years (over the entire dataset). We determine the same for papers that were not e-printed. The result is shown in figure 4.

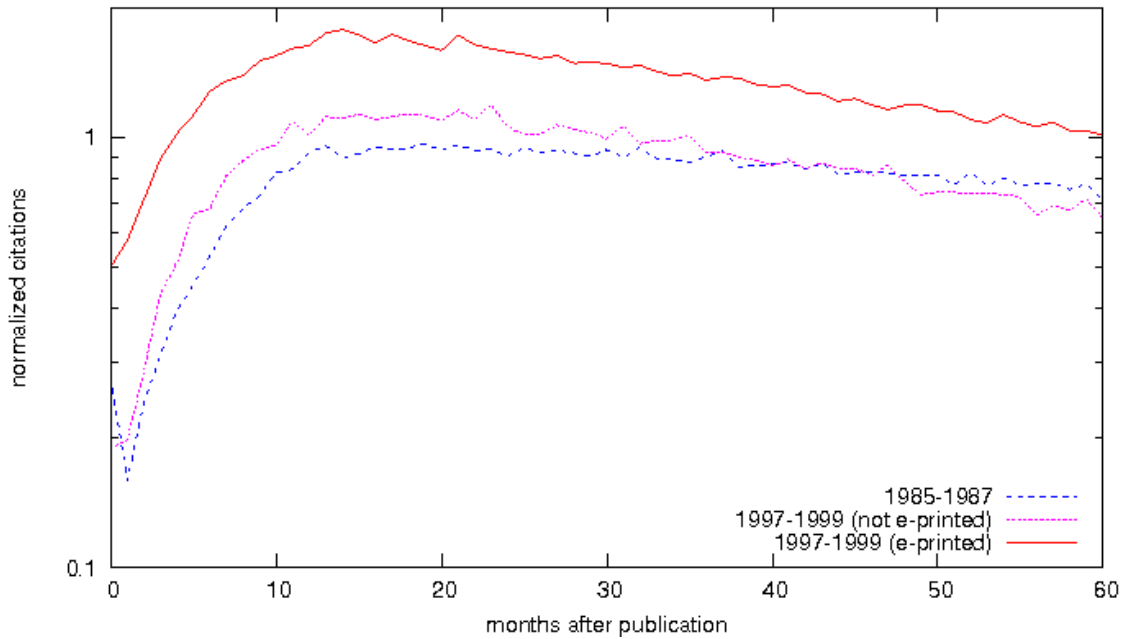

Figure 4. Normalized citation count for ApJ papers

Figure 4 shows that ApJ papers from the period 1997 through 1999 that were not e-printed follow the same citation trend as papers before the introduction of the arXiv. Just as in figure 3, figure 4 shows that the introduction of e-printing had a significant influence on citation rates. Our figure agrees with Schwartz and Kennicutt's observation (2004): "on average, ApJ papers posted on astro-ph are cited more than twice as often as those that are not posted on astro-ph". Figure 4 shows that we can extend this observation to ApJ papers from the pre-arXiv era.

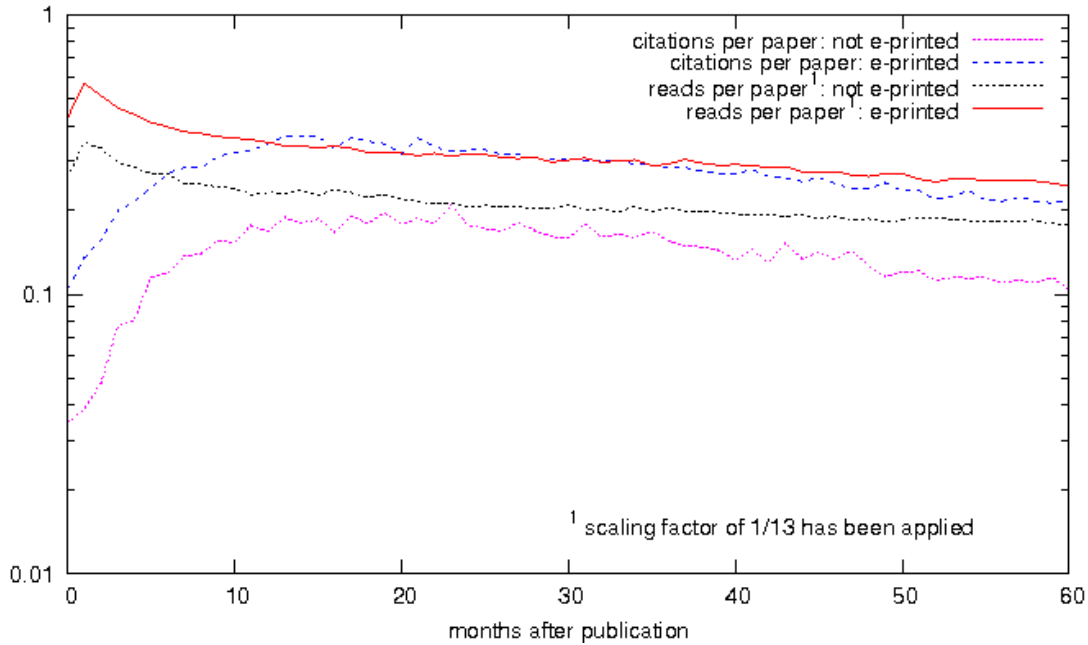

Figure 5. Citations and reads per paper for ApJ papers (1997-1999)

Figure 5 illustrates the correlation between "reads" (which stands for an "access" to the bibliographic record of a paper) and citations. E-printed ApJ papers are cited more and read more.

## 4. Discussion and Future Research

Since the introduction of the arXiv e-print repository, the percentage of e-printed papers has risen sharply for the astronomy and physics journals tracked in this paper (figure 1). When papers are sorted by month and citation count we take the top 100 of this list, we can see that this percentage rises even more sharply (figure 2). Figures 3 through 5 show significant increases in citation rates as a result of e-printing. Since the number of citations per paper two years after publication is closely related to the ISI Journal Impact Factor (Garfield 1972), figure 3 can be seen as the increase in the Journal Impact Factor after the introduction of e-printing. Figure 5 shows the strong correlation between "reads" and citations (see Kurtz et al. 2005b). Papers that appear as e-prints are both cited more and read more than papers that do not appear as e-prints. Our figures show that the effect of e-printing on citation rates in astronomy and physics is significant, which supports results obtained in the other studies cited.

We mentioned that Kurtz et al. (2005a) established that the higher citation rate of e-printed papers cannot be explained by assuming earlier availability (the EA postulate) as the sole reason. The authors' appraisals of the quality of the papers is an additional motivation to offer them to the arXiv repository prior to publication in a journal. The fact that we see higher citation rates for e-printed papers supports the Kurtz group's observation. This means that the best papers in astronomy appear as e-print first. If in a broad sense bibliometric trends in astronomy are typical for bibliometric trends in other disciplines, we feel that it is safe to say that the best physics papers also first appear as e-prints in the arXiv repository.

As indicated in Kurtz et al. (2005a), there might be additional effects contributing to increased citation rates (besides the three effects studied). For example, there may be a contribution of electronic publishing in general. The astronomy journals referenced in this paper introduced their electronic journals in the late 1990s, and the physics journals in the early 2000s. Although we do not have data in this study to support any evidence, we would expect that the higher probability of finding a paper using a search engine argues in favor of increased citation rates. For this same reason, there might be a contribution of sophisticated gateways to the online scholarly literature, like the ADS. Lawrence (2001) argues that articles available through open access sources have an increased readership (because of a higher probability of finding them), which leads to higher citation rates. On the other hand, putting back issues online for the core astronomy journals in the ADS did not increase citation rates to papers in those back issues. In addition, electronically published journals often still have high subscription fees, making them available only to well-funded disciplines. The influence of the electronic version of, say, the *Astrophysical Journal* on citation rates (if any), will probably be small compared to the contribution of the early access effect, and therefore difficult to detect. A researcher would have to study a journal without a significant e-printing culture that switched over to electronic publishing to fully analyze the influence. This falls outside the scope of this paper.

In future work we will look at additional impacts of e-printing, readership and citations of e-prints before and after the publication of the paper and we will use the complete ADS citation database for astronomy to analyze citation distributions.

**Credits**

The authors would like to thank Paul Ginsparg and Simeon Warner for providing data from the arXiv. The Astrophysics Data System is funded by NASA Grant NCC5-18.

## Appendix A - Journal Abbreviations Used

| Journal | Abbreviation |
| --- | --- |
| The Astrophysical Journal, Letters, Supplement Series | ApJ, ApJL, ApJS |
| The Astronomical Journal | AJ |
| Astronomy & Astrophysics | A+A |
| Monthly Notices of the Royal Astronomical Society | MNRAS |
| Publications of the Astronomical Society of the Pacific | PASP |
| The Physical Review A,B,… | PhRvA, PhRvB, … |
| Nuclear Physics B | NuPhB |
| Journal of Chemical Physics | JChPh |

## Authors' involvement in the ADS

**Edwin Henneken** has been working as a programmer for the ADS since 2002. He is responsible for developing and maintaining software used for harvesting and processing metadata, the extraction of metadata from PDF files and OCR results, and the processing of scans. Currently he is also using the ADS data logs and citation statistics for bibliometric research. He graduated in computational astrophysics at the Leiden Observatory in The Netherlands, did micro meteorological research on turbulent energy exchange over the Greenland ice sheet, co-founded a Web-design and technical-translation business, and worked in the Internet industry, managing a global roaming service. Your may reach him at ehenneken@cfa.harvard.edu

**Michael Kurtz** is an astronomer and computer scientist at the Harvard-Smithsonian Center for Astrophysics in Cambridge, Massachusetts, which he joined after receiving a Ph.D. in Physics from Dartmouth College in 1982. Kurtz is the author or co-author of over 200 technical articles and abstracts on subjects ranging from cosmology and extra-galactic astronomy, to data reduction and archiving techniques, to information systems and text retrieval algorithms. In 1988 Kurtz conceived what has now become the NASA Astrophysics Data System, the core of the digital library in astronomy, perhaps the most sophisticated discipline-centered library extant. He has been associated with the project since that time, and was awarded the 2001 Van Biesbroeck Prize of the American Astronomical Society for his efforts. You may reach him at kurtz@cfa.harvard.edu

**Guenther Eichhorn** is the Project Scientist for the Astrophysics Data System, a NASA funded project that makes the astronomical literature available on-line. He is employed by the Smithsonian Institution at the Harvard-Smithsonian Center for Astrophysics. By education he is an astronomer and has worked on diverse subjects like interplanetary dust detectors, age determinations of lunar samples, space based gamma ray detectors, and speckle interferometry. Since 1992 he has been leading the efforts to provide a comprehensive literature search system for Astronomy and Physics. The Special Libraries Association awarded him the 2001 Physics, Astronomy and Mathematics Division Award for his efforts in developing this system. You may reach him at gei@cfa.harvard.edu

**Carolyn Grant** has been working at the Harvard-Smithsonian Center for Astrophysics for over twenty years and has been programming for the ADS abstract service since its inception thirteen years ago. She originally began working at CfA while an undergraduate at Harvard College and liked it so much she never left. Her primary responsibilities include managing and updating the databases and coordinating with data suppliers. You may reach her at stern@cfa.harvard.edu

**Alberto Accomazzi** is a senior IT specialist at the Harvard-Smithsonian Center for Astrophysics, in Cambridge, MA. He currently works on developing and maintaining much of the software used to process and index the bibliographic databases that are part of the Astrophysics Data System. His professional interests are in the areas of natural language processing, distributed systems, classification, image processing, optical character recognition, and open source initiatives. You may reach him at aaccomazzi@cfa.harvard.edu

**Donna Thompson** is the librarian for the Smithsonian/NASA Astrophysics Data System. She has been with the ADS since 1998. In addition to Library Science, she has studied education and history. She gathers new data for inclusion in the various databases in the ADS and also coordinates the historical literature project. You may reach her at dthompson@cfa.harvard.edu.

**Stephen S. Murray** is Senior Astrophysicist and Deputy Director for Science at the Harvard-Smithsonian Center for Astrophysics. He is Principal Investigator for the ADS and for the High Resolution Camera for the Chandra X-ray Observatory. His main areas of interest are extragalactic astronomy, high resolution imaging instrumentation, and information systems. A member of the original group of young X-ray astronomers who joined the Smithsonian Astrophysical Observatory in the mid-1970s, he has been actively involved in what was called the Advanced X-ray Astrophysics Facility program (later renamed Chandra) since the initial proposals for an orbiting telescope were sent to NASA more than two decades ago. He has been involved with the ADS since 1988. You may reach him at ssm@cfa.harvard.edu